\documentclass[conference]{IEEEtran}
\IEEEoverridecommandlockouts
\usepackage{amsmath,amsfonts}
\usepackage{algorithmic}
\usepackage{array}
\usepackage[caption=false,font=footnotesize,labelfont=rm,textfont=rm]{subfig}
\usepackage{textcomp}
\usepackage{stfloats}
\usepackage{url}
\usepackage{verbatim}
\usepackage{graphicx}
\usepackage{balance}
\usepackage{xcolor}
\usepackage{epstopdf}
\usepackage{multirow}
\usepackage{booktabs}
\usepackage{latexsym}
\usepackage{cite}
\usepackage{algorithm}
\usepackage{hyperref}
\usepackage{marvosym}
\usepackage[numbers, sort&compress]{natbib}
\usepackage{nomencl}
\usepackage{textcomp,mathcomp}
\usepackage{color}
\usepackage{enumitem}
\usepackage{mathrsfs}

\newenvironment{termtable}[1][2cm]{%
	\def\term##1##2{\item[$##1$] ##2}%
	\itemize[left=0pt .. #1, itemindent=0pt,
	align=parleft, nosep]
}
{%
	\enditemize
}
\makenomenclature

\begin{document}
	
	\title{
		Aggregating Inverter-Based Resources for Fast Frequency Response: A Nash Bargaining Game-Based Approach
	}
	
	\author{\IEEEauthorblockN{Xiang Zhu, Hua Geng, Hongyang Qing}
		\IEEEauthorblockA{\textit{Department of Automation, Tsinghua University} \\
			Beijing 100084, China \\
			zhu-x22@mails.tsinghua.edu.cn, \\
            \{genghua, qinghy\}@mail.tsinghua.edu.cn
            }
		\and
		\IEEEauthorblockN{Xin Zou}
		\IEEEauthorblockA{\textit{State Grid Economic and Technological Research Institute Co., Ltd.} \\
			Beijing 102209, China \\
			zouxin@chinasperi.sgcc.com.cn}		
            
		\thanks{This work was supported by Science and Technology Project of State Grid Corporation of China (No. 5100-202456019A-1-1-ZN).~(\emph{Corresponding author: Hua Geng})
        }
	}
	
	\maketitle
	
	\begin{abstract}  	

        This paper proposes a multi-objective optimization~(MOO) approach for grid-level frequency regulation by aggregating inverter-based resources~(IBRs). Virtual power plants~(VPPs), acting as aggregators, can efficiently respond to dynamic response requirements from the grid. Through parametric modeling, grid-level frequency regulation requirements are accurately quantified and translated into a feasible parameter region defined by device-level parameters. Based on this feasible region, an MOO model is developed to address the conflicting demands of IBRs during frequency response. A Nash bargaining game-based approach is then employed to optimally allocate regulation requirements within the VPP, balancing the various demands of the IBRs. Numerical experiments demonstrate the effectiveness of the proposed method in enhancing frequency stability and improving coordination among IBRs.
		
	\end{abstract}
	
	
	\IEEEpeerreviewmaketitle
	\section*{Nomenclature}
	
	\subsection*{\bf{Abbreviations}}
	\begin{termtable}[1.9cm]
		\term{\text{SG}}{Synchronous generator}
        \term{\text{PV}}{Photovoltaic}
		\term{\text{IBR}}{Inverter-based resources}
        \term{\text{FFR}}{Fast frequency response}
        \term{\text{PFR}}{Primary frequency regulation}
        \term{\text{MOO}}{Multi-objective optimization}
		\term{\text{VPP}}{Virtual power plant}
        \term{\text{Qss}}{Quasi-steady state}
        \term{\text{RoCoF}}{Rate of the change of frequency}
	\end{termtable}
    
    \subsection*{\bf{Parameters}}
	\begin{termtable}[1.9cm]
		\term{\Delta P}{The amplitude of power disturbance [MW]}
		\term{D_{0}}{The load damping coefficient [MW/Hz]}
		\term{H_{0}}{The inertia parameter of SGs [MWs/Hz]}
		\term{R}{The droop coefficient of SGs [MW/Hz]}
		\term{T^\text{SG}}{The response time constant of SGs [s]}
		\term{f_\text{DB1}}{Width of dead band for VPP [Hz]}	
        \term{f_\text{DB2}}{Width of dead band for SGs [Hz]}	
		\term{\Delta f_\text{lim}^\text{RoCoF}}{RoCoF limitation of frequency [Hz/s]}
		\term{\Delta f_\text{lim}^\text{Nadir}}{Nadir limitation of frequency [Hz]}
		\term{\Delta f_\text{lim}^\text{Qss}}{Quasi-steady-state limitation of frequency [Hz]}
		\term{N}{The number of IBRs involved in the VPP}
	\end{termtable}

    \subsection*{\bf{Variables}}
    \begin{termtable}[1.9cm]
        \term{\Delta f_\text{max}^\text{RoCoF}}{RoCoF metric [Hz/s]}
		\term{\Delta f^\text{Nadir}}{Nadir metric[Hz]}
		\term{\Delta f^\text{Qss}}{Quasi-steady-state metric[Hz]}
        \term{H_{k}}{Virtual inertia of the $k$th IBR [MWs/Hz]}
		\term{D_{k}}{Virtual damping of the $k$th IBR [MW/Hz]}
    \end{termtable}
	
	\section{Introduction}
	
	\IEEEPARstart{I}{n} future power systems, traditional synchronous generators~(SGs) with substantial inertia are steadily being replaced by inverter-based resources~(IBRs). This transition, caused by the reduction of SGs, significantly decreases system inertia, necessitating the use of IBRs to provide ancillary services, such as inertia support and primary frequency response~\cite{he2022transient}.
	
	Virtual power plants~(VPPs) play a crucial role in enhancing frequency regulation capabilities in the low-inertia power system~\cite{VPP-ruan}. By aggregating and coordinating numerous IBRs through business contracts, VPPs can operate as unified entities that dynamically respond to grid requirements. This collective approach unlocks the potential of fast frequency response from IBRs so as to enable more efficient frequency regulation and guarantee the frequency security of the grid~\cite{VPP2,SFR1}.

	A primary challenge in grid frequency regulation lies in optimizing the output of inverter-based resources (IBRs) to meet the grid's frequency regulation requirements~\cite{DER, feng2025hybrid}. Ensuring frequency security requires maintaining both the rate of frequency change and the frequency deviation (including dynamic and steady-state conditions) within safe operational limits. Previous studies have proposed various approaches to address these challenges. For instance, the virtual inertia scheduling (VIS) scheme was utilized in~\cite{VIS} to optimize IBR output and ensure frequency security. Additionally, the dual-mode operation of IBRs for output modulation was explored in~\cite{FFR3}. In~\cite{COM2}, a frequency security-constrained stochastic look-ahead scheduling approach was introduced to enhance the online modulation of IBR power injections. Moreover, a minimal reserve decision and optimal allocation scheme were proposed in~\cite{zhu2025optimal} to account for time-varying active power injections, thereby eliminating idle reserve for IBRs. Despite these advancements, existing approaches often overlook the diverse operational characteristics and requirements of individual IBRs, which often conflict with one another. This gap presents a significant challenge in coordinating IBRs for grid-level frequency regulation.

    When allocating frequency regulation responsibilities within a VPP, the inherent conflicts among inverter-based resources (IBRs) introduce significant complexity in optimizing their output. Traditional single-objective optimization approaches~\cite{VIS, FFR3, COM2, zhu2025optimal} are insufficient in addressing the diverse operational priorities and constraints of individual IBRs within a VPP. These approaches often fail to account for the competing and sometimes conflicting requirements of the various resources. To overcome these limitations, multi-objective optimization (MOO) methods are increasingly employed to handle the multiple, and often conflicting, demands~\cite{Zhu-TIA}. By utilizing MOO, a VPP can effectively balance the internal resource demands, ensuring that each IBR operates optimally while simultaneously fulfilling the grid-level ancillary service needs, thereby achieving a more reasonable and feasible approach to frequency regulation.

    In this paper, we propose an MOO approach for VPPs to provide effective grid ancillary services. Through parameterized modeling, the frequency regulation requirements are accurately quantified and represented as a closed-form feasible region for the device-level control parameters. By incorporating this feasible region into the optimization process, we develop an MOO model that facilitates the effective coordination of the diverse needs of various IBRs. This approach addresses the inherent conflicts among IBRs and supports efficient frequency regulation, ensuring optimal performance in meeting grid ancillary service requirements.

	The major contributions of this paper are twofold:
	\begin{enumerate}
		
		\item A parametric modeling approach is proposed to derive the closed-form frequency regulation requirements for IBRs. Both the security metrics and recovery rate of system frequency are considered to formulate the feasible parameter region for the required virtual inertia and damping parameters.
		
		\item An MOO model is developed to allocate the required virtual inertia and damping to IBRs, taking into account both the economic benefits of the VPP and the diverse demands of IBRs. A Nash bargaining-based approach is introduced to determine the optimal strategy that balances these conflicting objectives.
		
	\end{enumerate}
	
	\section{Problem Formulation}
    
\begin{figure}
		\centering
		\includegraphics[width=1\linewidth]{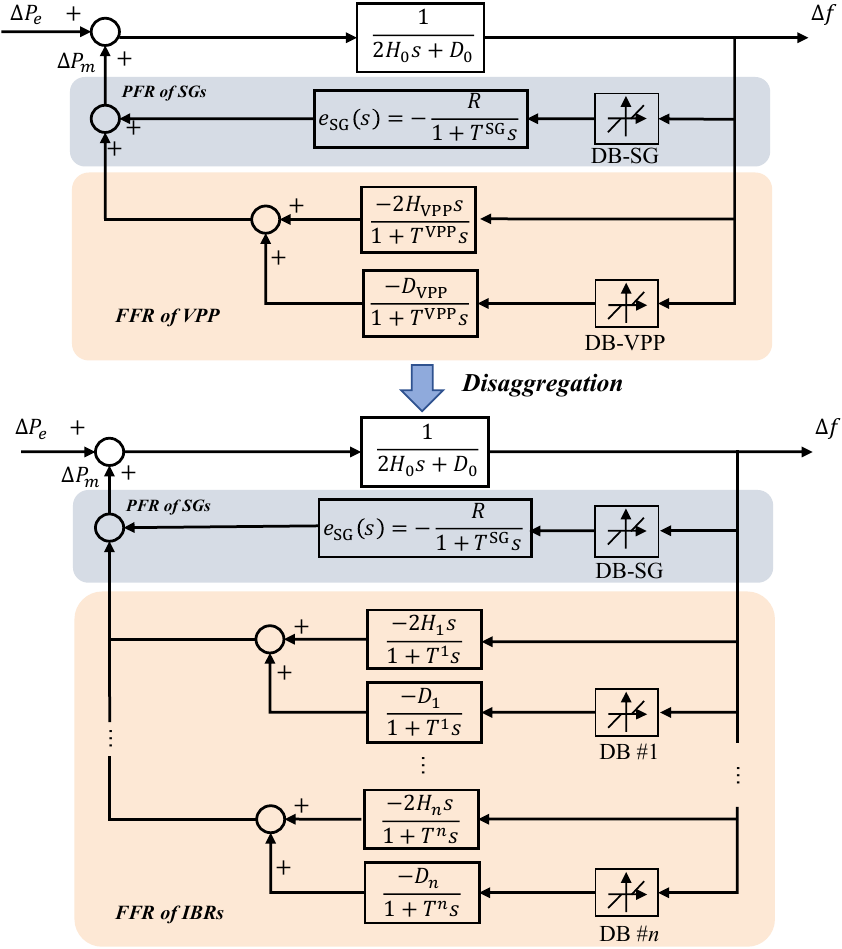}
		\caption{The block diagram illustrates the system frequency response. The forward path represents the inertia and damping of the grid. While the orange part and gray part describe the frequency responses of SGs and VPP, respectively.} 
		\label{fig.block}
	\end{figure}

	In this paper, we consider a VPP containing various controllable IBRs, i.e., solar panels and wind turbines, to provide grid ancillary services. As for the frequency regulation services, the VPP aggregates IBRs working on virtual inertia and droop control scheme to provide the required active power injections~\cite{FFR2}. In Fig.~\ref{fig.block}, a block diagram is used to describe the relationship between the system frequency deviation ($\Delta f$) and power disturbance ($\Delta P_{e}$) based on the swing equation \cite{COM2}. Three feedback loops, i.e., the primary frequency regulation of SGs, the inertia response and the primary frequency regulation of VPP, are added to represent the active power injections of SGs and VPP. Besides, dead bands are considered in the feedback loops of PFR to reduce the operation and maintenance costs~\cite{DB2}.  
	
    The frequency response performance of an IBR is determined by its active power injection dynamics, which satisfy the swing equation of power-frequency. The active power injection of the IBR comes from the regulation power reserve that is previously prepared to support frequency regulation~\cite{COM2}. To incentivize IBRs to actively participate in system frequency regulation, the power grid typically offers economic compensation based on the actual power injection of the IBR. Consequently, once the power reserve of each IBR is determined, the allocation of actual power injection is closely tied to the economic benefits of each IBR, often leading to conflicting interests.

    As the aggregator, the VPP should strategically allocate the regulation power injection across IBRs, ensuring that each IBR has an acceptable level of reserve utilization. This allocation process needs to balance the economic interests of individual IBRs while simultaneously ensuring the requirements of system frequency security.
    
	\section{Parametric Modeling of Frequency and Requirements Derivation}
	
	When an active power imbalance occurs, the power system frequency deviates from its nominal value, reaching a peak deviation before gradually settling toward a quasi-steady-state value. Based on this dynamic behavior, the regulation requirements can be defined as maintaining the frequency within a secure range. The specific requirements associated with this regulation are discussed in detail in this section.

    \subsection{Frequency Dynamics Modeling}
    
    For simplification, the dead bands for VPP and SGs are represented by $f_\text{DB1}/s$ and $f_\text{DB2}/s$ superimposing to the frequency deviation, and we assume $f_\text{DB1} < f_\text{DB2}$. Besides, considering the response time constant of SGs~($T^\text{SG}$) is larger than those of VPP and IBRs, therefore the response time constants $T^\text{VPP}$, $T^{1}$, ..., $T^{n}$ could be omitted.
    
	Without loss of generality, we consider the frequency drop and derive the system frequency response function in Eq.~(\ref{fre_2}). The input, $\Delta P_{e}(s)$, is modeled as a step disturbance $-\Delta P/s$, the parameter $R$ is the aggregated droop coefficient of SGs, $H_{0}$ is the aggregated inertia parameter of SGs, $D_{0}$ is the load damping coefficient. The parameters $H_\text{VPP}$ and $D_\text{VPP}$ represent the virtual inertia and virtual damping of the VPP, reflecting the aggregated characteristics of the IBRs. Additionally, $t_\text{DB1}$ and $t_\text{DB2}$ are the time points when the PFR of VPP and SGs are activated, respectively. Due to the very short time interval between these two events, the frequency response between $t_\text{DB1}$ and $t_\text{DB2}$ is considered negligible and is therefore omitted.
	
	\begin{equation}
		\left\{ \begin{aligned}
			&  \Delta f(t) = - \frac{{\Delta P}}{{{D_0}}} \cdot \left( {1 - {e^{ - \frac{{{D_0}}}{{2H}} \cdot t}}} \right),0 \le t \le {t_\text{DB1}}\\ 
			&  \Delta f(t) = - \frac{{\Delta P + {D_\text{VPP}}{f_\text{DB1}}}}{{{D_\text{VPP}} + {D_0} + R}} \left[ {1 + {e^{ - \zeta {\omega _n}t}}{\eta _1}\sin ({\omega _d}t + {\varphi _1})} \right] \\
			& - \frac{{R \cdot {f_\text{DB2}}}}{{{D_\text{VPP}} + {D_0} + R}} \left[ {1 - {e^{ - \zeta {\omega _n}t}}{\eta _2}\sin ({\omega _d}t + {\varphi _2})} \right],t \ge {t_\text{DB2}}\\ 
		\end{aligned} \right.
		\label{fre_2}
	\end{equation}
	where
	\begin{subequations}
		\begin{align}
			& {{\omega }_{n}}=\sqrt{\frac{D+R}{2H{{T}^\text{SG}}}},\zeta =\frac{2H+D{{T}^\text{SG}}}{2\sqrt{2{{T}^\text{SG}}H(R+D)}}, \\
			& {{\omega }_{d}}={{\omega }_{n}}\sqrt{1-{{\zeta }^{2}}},\eta_{1} =\sqrt{\frac{1-2T^\text{SG}{{\omega }_{n}}\zeta +{{T^\text{SG}}^{2}}\omega _{n}^{2}}{1-{{\zeta }^{2}}}}, \\
			&  \varphi_{1} =\arctan (\frac{{{\omega }_{d}}}{-T^\text{SG}\omega _{n}^{2}+\zeta {{\omega }_{n}}}),\eta_{2}=\frac{1}{\sqrt{1-{{\zeta }^{2}}}}, \\
			& \varphi_{2}=\arctan (\frac{\sqrt{1-{{\zeta }^{2}}}}{\zeta }),H = H_{0}+H_\text{VPP}, D = D_{0}+D_\text{VPP}
		\end{align}	
	\end{subequations}

	\subsection{Frequency Security Constraints}
	
	In the deviation stage, the objective is to maintain a safe rate of the change of frequency~(RoCoF) as well as a maximum deviation value called the nadir value of the system frequency. Based on the time-domain representation of the frequency, the maximal RoCoF is derived in Eq.~(\ref{con.1}), which refers to the RoCoF at the beginning of deviation. 
	\begin{equation}
		\Delta f_{\max }^\text{RoCoF}={{\left. \Delta \dot{f}(t) \right|}_{t={{0}^{+}}}}=-\frac{\Delta P}{2(H_{0}+H_\text{VPP})}
		\label{con.1}
	\end{equation}
	
	Similarly, the nadir value of the frequency is derived in Eq.~(\ref{con.2}) by solving its extreme point. 
	\begin{equation}
		\begin{aligned}
			{\left. \Delta {{f}^\text{Nadir}} \right|}_{\Delta \dot{f}(t_{n})=0} = \Delta f({{t}_{n}})
		\end{aligned}	
		\label{con.2}
	\end{equation}
	where $t_{n}$ is the time that frequency deviation reaches its lowest point (nadir value), which is derived in Eq.~(\ref{nadir1}).
	\begin{subequations}
		\begin{align}
			& {{t}_{n}}=\left(\arctan (N)\right)/{\omega }_{d}\\
			& N =\frac{{{\omega }_{d}}(m\cos ({\varphi_{2}})-\cos (\varphi_{1}))-\zeta {{\omega }_{n}}(m\sin ({\varphi_{2}})+\sin (\varphi_{1}))}{\zeta {{\omega }_{n}}(m\cos({\varphi_{2}})-\cos (\varphi_{1}))+{{\omega }_{d}}(m\sin ({\varphi_{2}})-\sin (\varphi_{1}))} \\
			& m=\frac{R{{f}_\text{DB2}}}{\Delta P+{{D}_\text{VPP}}{{f}_\text{DB1}}}\cdot \frac{{{\eta_{2}}}}{\eta_{1}}
		\end{align}
		\label{nadir1}
	\end{subequations}
	
	Besides, based on the time-domain representation of the frequency, the quasi-steady-state~(Qss) value is derived in Eq.~(\ref{con.3}) and sets a constraint on virtual damping.
	\begin{equation}
		\Delta {{f}^\text{Qss}}={{\left. \Delta f(t) \right|}_{t\to \infty }}=-\frac{\Delta P+{{D}_\text{VPP}}{{f}_\text{DB1}}+R\cdot {{f}_\text{DB2}}}{{{D}_\text{VPP}}+{{D}_{0}}+R}
		\label{con.3}
	\end{equation}	
	
	To sum up, the frequency regulation requirements of three security metrics are derived in Eq.~(\ref{safety-1}), which are mapped to the parameter region of virtual inertia ($H_\text{VPP}$) and damping ($D_\text{VPP}$).
	
	\begin{equation}
		\mathcal{F} = \left\{ \begin{aligned}
			&  H_\text{VPP} \ge \frac{\Delta P}{2{\Delta f_{\lim }^\text{RoCoF}}}-H_{0} \\ 
			&  \left| \Delta f({{t}_{n}}) \right| \le \Delta f_{\lim }^\text{Nadir}\\ 
			& D_\text{VPP} \ge \frac{\Delta P+R f_\text{DB2}-\Delta f_{\lim }^\text{Qss}(D_{0}+R)}{\Delta f_{\lim }^\text{Qss}-f_\text{DB1}}\\
			& 0 \le H_\text{VPP} \le H_\text{VPP}^{\max}, 0 \le D_\text{VPP} \le D_\text{VPP}^{\max}\\
		\end{aligned} \right.
		\label{safety-1}
	\end{equation}
	
	\subsection{Required Inertia and Damping Determination}

    The scheduling scheme for active power injections from IBRs is typically implemented through a two-phase process: virtual inertia and damping parameter determination, followed by their optimal allocation within the VPP framework. On the basis of our earlier work~\cite{zhu2025optimal}, the required inertia ($H_\text{VPP}^\text{re}$) and damping ($D_\text{VPP}^\text{re}$) can be determined by a two-stage sequential algorithm with in the derived $\mathcal{F}$ in Eq.~(\ref{safety-1}) that ensure the security of system frequency. 
	
	\section{Multi-Objective Optimization for Requirements Allocation}
	
	\subsection{Objective Function of the VPP}
	
	\begin{figure}
		\centering
		\includegraphics[width=1\linewidth]{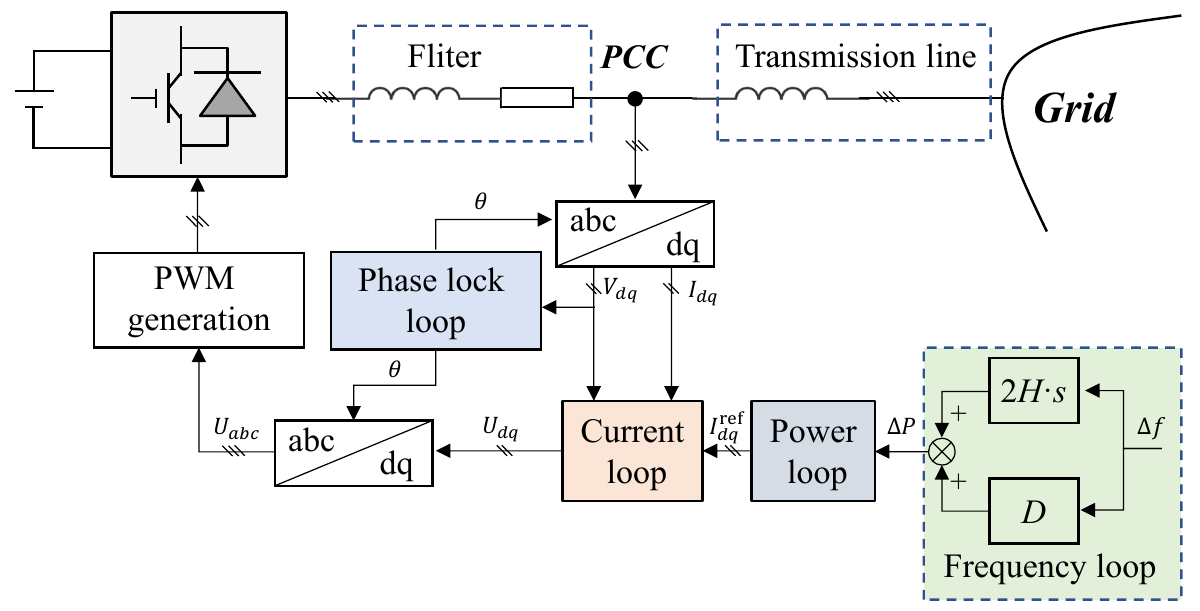}
		\caption{The control diagram of inverter-based resources, which contains a frequency control loop dominated by the virtual inertia ($H$) and damping ($D$) parameters.} 
		\label{fig.IBR}
	\end{figure}
	
	Based on the determined values of $H_\text{VPP}^\text{re}$ and $D_\text{VPP}^\text{re}$, the regulation requirements for the VPP are defined and should be fully allocated to the IBRs. By applying the principle of linear superposition, the virtual inertia and damping requirements for the VPP can be distributed among the control loop parameters of the IBRs equivalently, such that $H_\text{VPP}^\text{re} = \sum_{i=1}^{N} H_{k}$ and $D_\text{VPP}^\text{re} = \sum_{i=1}^{N} D_{k}$. Each IBR then adjusts its active power injection according to the allocated virtual inertia and damping parameters, as illustrated in Fig.~\ref{fig.IBR}.	
	
	Thus, the frequency support provided by the IBRs can be explicitly quantified through a mapping approach using the parameter combination ($H_{k}$, $D_{k}$). Inspired by the approach in~\cite{VIS}, different allocation schemes of ($H_\text{VPP}^\text{re}$, $D_\text{VPP}^\text{re}$) lead to varying regulation costs due to the differing economic preferences of the IBRs. Based on this observation, the objective function of the VPP can be constructed as follows.
	\begin{equation}
		\max \left[ {a H_{{\rm{VPP}}}^\text{re} + b D_{{\rm{VPP}}}^\text{re} - \left( {\sum\nolimits_{k = 1}^N {{\alpha _k}{H_k} + {\beta _k}{D_k}} } \right)} \right]
		\label{obj-1}
	\end{equation}
	where $\left(a H_{{\rm{VPP}}}^{re} + b D_{{\rm{VPP}}}^{re}\right)$ is the economic compensation for the VPP from power grid while $\left( {\sum\nolimits_{k = 1}^N {{\alpha _k}{H_k} + {\beta _k}{D_k}} } \right)$ represents the regulation cost of the VPP which is dominated by the economic variety of IBRs.
	
	Due to the first term of Eq.~(\ref{obj-1}) can be seen as a constant value under a frequency regulation period, Eq.~(\ref{obj-1}) can be transferred into Eq.~(\ref{obj-2}) equivalently.
	\begin{equation}
		\min C_\text{VPP} = \left( {\sum\nolimits_{k = 1}^N {{\alpha _k}{H_k} + {\beta _k}{D_k}} } \right)
		\label{obj-2}
	\end{equation}

    \subsection{Multi-Objective Optimization Model Formulation}
    
	For the IBRs, the goal is to maximize their returns within the permissible active power injection range for frequency regulation. Accordingly, the set of objective functions for the IBRs is formulated as shown in Eq.~(\ref{obj-3}).
	
	\begin{equation}
		\max \left\{ {\frac{{\Delta {P_1}}}{{{P_{a,1}}}},\frac{{\Delta {P_2}}}{{{P_{a,2}}}},...,\frac{{\Delta {P_N}}}{{{P_{a,N}}}}} \right\}
		\label{obj-3}
	\end{equation}
	where $\Delta P_{k}(k=1,2,...,N)$ and $P_{a,k}(k=1,2,...,N)$ represent the actual energy injection as well as the maximum energy injection of the $k$th IBR for frequency regulation, respectively. Thus, the ratio $\frac{\Delta P_{k}}{P_{a,k}}(k=1,2,...,N)$ describes the power reserve utilization for frequency support. 
	
	With a relatively short transient period, the active power injection is approximately in direct proportion to the damping parameter. Therefore, the objection functions in Eq.~(\ref{obj-3}) can be transferred in Eq.~(\ref{obj-4}). 
	\begin{equation}
		\min \left\{ {(D_{{\rm{VPP}}}^\text{re}\frac{{{P_{a,1}}}}{{\Delta P}} - {D_1}),...,(D_{{\rm{VPP}}}^\text{re}\frac{{{P_{a,N}}}}{{\Delta P}} - {D_N})} \right\}
		\label{obj-4}
	\end{equation}
	where $D_\text{VPP}^\text{re}\frac{P_{a,k}}{\Delta P}$ means the ideal damping being allocated for the $k$th IBR which maximizes its reserve utilization.	
	
	To sum up, the multi-objective optimization model contains (\ref{obj-2})-(\ref{obj-4}) is constructed as in Eq.~(\ref{opt-model}).
	
	\begin{equation}
		\begin{aligned}
			& \min \left\{ {\sum\nolimits_{k = 1}^N {\left( {{\alpha _k}{H_k} + {\beta _k}{D_k}} \right)} ,...,(D_{{\rm{VPP}}}^\text{re}\frac{{{P_{a,k}}}}{{\Delta P}} - {D_k}),...} \right\}\\
			& \text{s.t}\\
			& \sum\nolimits_k^N {{H_k}}  = {H^\text{re}_{{\rm{VPP}}}},\\
			& \sum\nolimits_k^N {{D_k}}  = {D^\text{re}_{{\rm{VPP}}}},\\
			& H_k^{\min } \le {H_k} \le H_k^{\max },\forall k \in \left\{ {1,2,...,N} \right\}\\
			& D_k^{\min } \le {D_k} \le D_k^{\max },\forall k \in \left\{ {1,2,...,N} \right\}
		\end{aligned}
		\label{opt-model}
	\end{equation} 
	
	

	\begin{figure}
		\centering
		\includegraphics[width=1\linewidth]{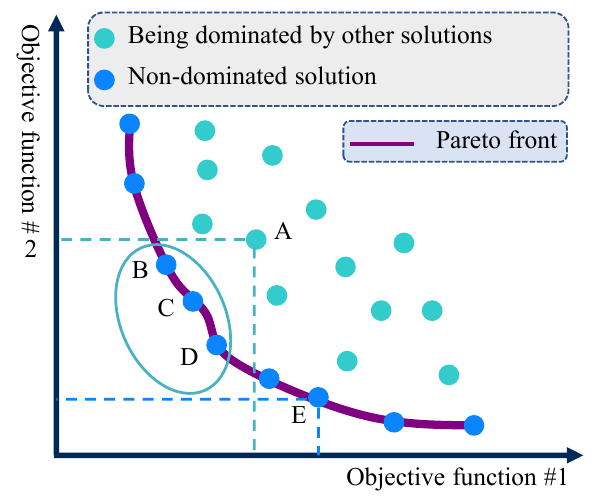}
		\caption{The diagram illustrates the Pareto front for a two-objective optimization problem. Solution A is not part of the Pareto front because there are existing solutions—B, C, and D—that offer better performance in both objectives. In contrast, solution E is a non-dominated solution and thus lies on the Pareto front, as no other solution improves both objectives simultaneously.} 
		\label{fig.pareto}
	\end{figure}    

	\begin{figure}
		\centering
		\includegraphics[width=1\linewidth]{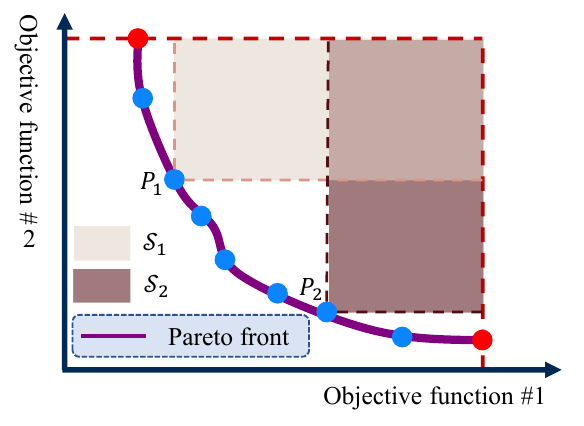}
		\caption{The diagram illustrates the bargaining criteria for a two-objective optimization problem where $\mathcal{S_\text{1}}$ for solution $P_{1}$ and $\mathcal{S_\text{2}}$ for solution $P_{2}$.} 
		\label{fig.bargain}
	\end{figure}    
    
	\subsection{Nash Bargaining-based Solution}
	
	For the MOO model in Eq.~(\ref{opt-model}), it is challenging to determine an exact strategy that satisfies all objectives due to inevitable conflicts between them. Inspired by the Nash Bargaining Game theory and its applications in power systems~\cite{Li-Nash,Wei-Nash}, we develop a Nash bargaining-based approach to identify trade-off solutions.

    The core of this approach involves first computing the Pareto front of Eq.~(\ref{opt-model}), followed by determining an appropriate bargaining solution. To generate the Pareto front, we vary the weight vector ($\lambda$) and minimize the multi-objective function defined in Eq.~(\ref{opt-eqmodel}), subject to the normalization constraint $\sum_{i=0}^{N}\lambda_{i}=1$. 
	\begin{equation}
		{\lambda_{0}\sum\nolimits_{k = 1}^N {\left( {{\alpha _k}{H_k} + {\beta _k}{D_k}} \right)} +...+\lambda_{N}(D_{{\rm{VPP}}}^{re}\frac{{{P_{a,N}}}}{{\Delta P}} - {D_N})} 
		\label{opt-eqmodel}
	\end{equation} 

    Fig.~\ref{fig.pareto} illustrates a schematic representation of the Pareto front for a two-objective optimization problem. The Pareto front consists of non-dominated solutions, meaning that no objective can be improved without causing a degradation in at least one of the other objectives. This front represents the trade-off surface, where each point reflects a different balance between the objectives. The final bargaining solution is selected from this set based on additional criteria that reflect the desired trade-offs.

    The bargaining criterion is defined by the area formed between the Pareto solution and the worst outcomes of the two objectives, as shown in Fig.~\ref{fig.bargain}. In this figure, two areas are illustrated: $\mathcal{S_\text{1}}$ for solution $P_{1}$ and $\mathcal{S_\text{2}}$ for solution $P_{2}$. The final bargaining solution is the solution with largest area, i.e., $\max \{\mathcal{S_\text{1}},\mathcal{S_\text{2}},...,\mathcal{S_\text{P}}\}$, where $P$ represents the set of all potential solutions. his approach ensures that the solution selected not only lies on the Pareto front but also maximizes the overall trade-off area, reflecting the best possible balance of the conflicting objectives.
    
	For simplicity, we denote the different objective functions as $F_\text{VPP}, F_{\text{IBR},1},...,F_{\text{IBR},N}$, and the set of weights as $\Omega = {\lambda_{0},...,\lambda_{N}}$. Under a specific $\Omega$, the Pareto solution of (\ref{opt-model}) can be represented as $F_{{\rm{VPP}}}^*,F_{{\rm{IBR,1}}}^*,...,F_{{\rm{IBR}},N}^*$. The bargaining solution for the original MOO model in Eq.~(\ref{opt-model}) is then obtained by maximizing the objective in Eq.~(\ref{opt-bargain}).
	\begin{equation}
		\left( {F_{{\rm{VPP}}}^u - F_{{\rm{VPP}}}^*} \right)\left( {F_{{\rm{IBR,1}}}^u - F_{{\rm{IBR,1}}}^*} \right) \cdots \left( {F_{{\rm{IBR}},N}^u - F_{{\rm{IBR}},N}^*} \right)
		\label{opt-bargain}
	\end{equation}
	where $F_{{\rm{VPP}}}^u, F_{{\rm{IBR,1}}}^u,..., F_{{\rm{IBR,N}}}^u$ are disagreement points for the VPP and numerous IBRs.	
	
	Based on this analysis, the Nash bargaining-based strategy for allocating frequency regulation requirements is a hierarchical optimization process. This involves computing the Pareto front by varying the weights for the objectives in Eq.~(\ref{opt-eqmodel}), followed by selecting the strategy that yields the highest value in Eq.~(\ref{opt-bargain}).
	
	\section{Case Study}
	
	\subsection{Simulation Setup}

	To illustrate the effectiveness of our work, the case studies focus on the frequency-drop event, which is triggered by an active power imbalance ($\Delta P < 0$) and $f_{0} = 50$ (Hz), $f_\text{DB1} = 0.03$ (Hz), $f_\text{DB2} = 0.033$ (Hz). As for the safety constraints of system frequency, the limit of RoCoF $\Delta f_\text{lim}^\text{RoCoF}$ is set to $0.4$ (Hz/s), the limit of nadir $\Delta f_\text{lim}^\text{Nadir}$ is set to $0.5$ (Hz), and the limit of Qss $\Delta f_\text{lim}^\text{Qss}$ is set to $0.35$ (Hz). The other detailed parameters~\cite{VIS} are listed in TABLE \ref{tab:parameter} and the simulation platforms are MATLAB R2022b and Gurobi 11.0.3 on a desktop with an Intel Core i7-10700 2.90GHz CPU. 
	
	\begin{table}[htbp]
		\centering
		\caption{System parameters for case studies}
		\begin{tabular}{ccc}
			\toprule
			Parameter & Value & Unit \\
			\midrule
			$D_{0}$ & 2     & p.u. \\
			$H_{0}$ & 10     & s \\
			$R$ & 25    & \textbackslash{} \\
			$T^\text{SG}$ & 5     & s \\
			$N$ & 8     & \textbackslash{} \\
			$\Delta P_{e}$ & 0.25   & p.u. \\
			$\alpha_{k}$ & 3, 4, 1, 1, 2, 1, 1, 1   & \$/s \\
			$\beta_{k}$ & 2, 3, 1, 1, 1.5, 1, 1, 1   & \$/p.u. \\
			\multirow{2}[0]{*}{$P_{k}^\text{rated}$} & 0.13, 0.1, 0.04, 0.05, & \multirow{2}[0]{*}{p.u.} \\
			& 0.05, 0.1, 0.02, 0.01 &  \\
			$H^\text{min}_{k}$ & 0.1  & s \\
			$D^\text{min}_{k}$ & 0.1  & p.u. \\
			\bottomrule
		\end{tabular}%
		\label{tab:parameter}%
	\end{table}%
	
	\subsection{Verification of Parametric Modeling}
	The frequency regulation requirements of the set frequency-drop events are derived based on the mapping modeling of the feasible region ($\mathcal{F}$). The required inertia and damping are computed as $H_\text{VPP}^\text{re}=19.125$~(s) and $D_\text{VPP}^\text{re}=12.109$~(p.u.). The corresponding security metrics are $\Delta f_{\max }^\text{RoCoF}=0.21$~(Hz/s), $\Delta {{f}^\text{Nadir}}=0.5$~(Hz) and $\Delta {{f}^\text{Qss}}=0.35$~(Hz), which satisfy the presupposed demands.
	
	\subsection{Nash Bargaining-based Strategy for Allocation}	
    
\begin{figure}
		\centering
		\includegraphics[width=1\linewidth]{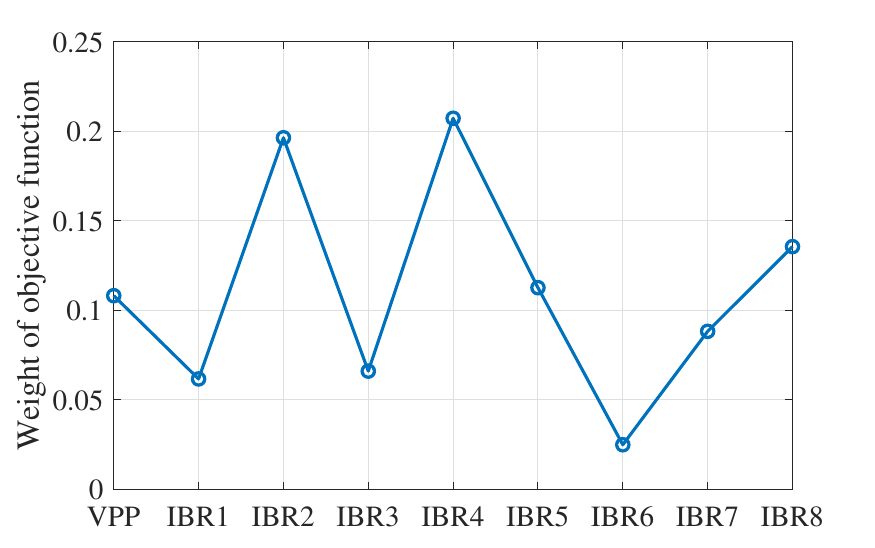}
		\caption{The optimal weights of objective functions of VPP and IBRs.} 
		\label{fig.weight}
	\end{figure}

	The Nash bargaining solution for the MOO model (\ref{opt-model}) is obtained through random combinations of weights, $\Omega = {\lambda_{0},...,\lambda_{N}}$, with 200 combinations selected. After solving the Pareto front and selecting the Nash bargaining solutions, the optimal weights for the VPP and eight IBRs are solved and shown in Fig. \ref{fig.weight}. The results reveal significant heterogeneity in weights among the nine participants, demonstrating their technical diversity and negotiated compromise. Notably, while all Pareto-optimal solutions satisfy feasibility constraints, the identified weights correspond to configurations achieving optimal VPP-IBRs coordination.
    
\begin{figure}
		\centering
		\includegraphics[width=1\linewidth]{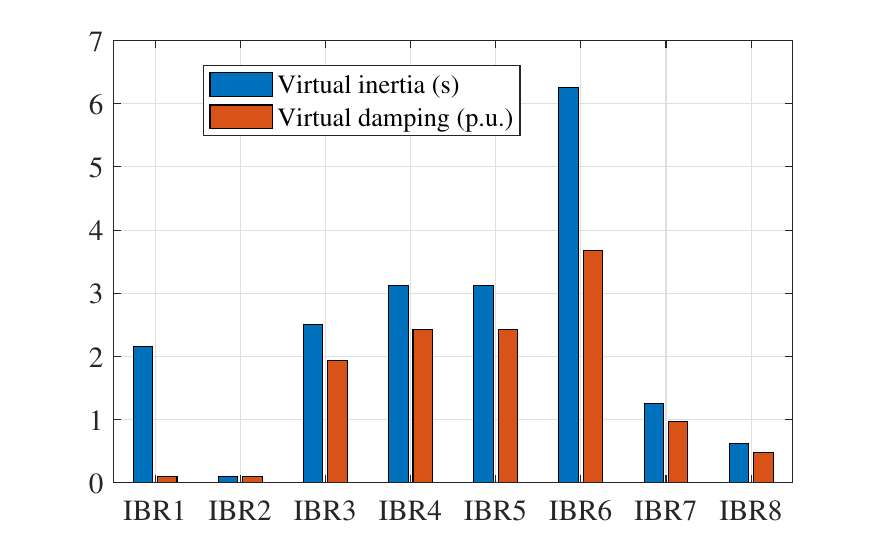}
		\caption{The optimal required inertia and damping allocation for eight IBRs.} 
		\label{fig.alloc}
	\end{figure}

    The corresponding allocation schemes of required virtual inertia and damping are displayed in Fig. \ref{fig.alloc}. The parameter allocation scheme reflects the economic preference of eight IBRs. Compared to the single-objective optimization focused solely on the economic benefit of the VPP, the proposed method increases the Nash bargaining value by 29.88\%, while reducing the VPP's economic benefit by 1.47\%. This trade-off demonstrates the effectiveness of the proposed approach in balancing economic efficiency with the fair distribution of virtual inertia and damping, thereby enhancing the collaborative performance of the IBRs within the VPP.
	
	\section{Conclusion}
	
	This paper proposes an MOO approach for VPPs aggregating IBRs to provide effective frequency regulation. The developed model introduces a Nash bargaining game to coordinate IBRs under constraints of system frequency security metrics. This methodology effectively resolves conflicts among heterogeneous IBRs while achieving Pareto-optimal equilibrium solutions. Case studies validate the effectiveness of the proposed approach, demonstrating a 29.88\% improvement in bargaining value and a 1.47\% reduction in economic costs compared to the single economic-driven method.

    Future work will focus on developing a unified MOO architecture capable of coordinating heterogeneous IBRs with diverse operational requirements under reduced computational complexity.  
	
	\footnotesize
	\bibliography{reference}

\begin{thebibliography}{10}
\providecommand{\url}[1]{#1}
\csname url@samestyle\endcsname
\providecommand{\newblock}{\relax}
\providecommand{\bibinfo}[2]{#2}
\providecommand{\BIBentrySTDinterwordspacing}{\spaceskip=0pt\relax}
\providecommand{\BIBentryALTinterwordstretchfactor}{4}
\providecommand{\BIBentryALTinterwordspacing}{\spaceskip=\fontdimen2\font plus
\BIBentryALTinterwordstretchfactor\fontdimen3\font minus
  \fontdimen4\font\relax}
\providecommand{\BIBforeignlanguage}[2]{{%
\expandafter\ifx\csname l@#1\endcsname\relax
\typeout{** WARNING: IEEEtran.bst: No hyphenation pattern has been}%
\typeout{** loaded for the language `#1'. Using the pattern for}%
\typeout{** the default language instead.}%
\else
\language=\csname l@#1\endcsname
\fi
#2}}
\providecommand{\BIBdecl}{\relax}
\BIBdecl

\bibitem{he2022transient}
C.~He, X.~He, H.~Geng, H.~Sun, and S.~Xu, ``Transient stability of low-inertia
  power systems with inverter-based generation,'' \emph{IEEE Transactions on
  Energy Conversion}, vol.~37, no.~4, pp. 2903--2912, 2022.

\bibitem{VPP-ruan}
G.~Ruan, D.~Qiu, S.~Sivaranjani, A.~S. Awad, and G.~Strbac, ``Data-driven
  energy management of virtual power plants: A review,'' \emph{Advances in
  Applied Energy}, vol.~14, p. 100170, 2024.

\bibitem{VPP2}
V.~Häberle, A.~Tayyebi, X.~He, E.~Prieto-Araujo, and F.~Dörfler,
  ``Grid-forming and spatially distributed control design of dynamic virtual
  power plants,'' \emph{IEEE Transactions on Smart Grid}, vol.~15, no.~2, pp.
  1761--1777, 2024.

\bibitem{SFR1}
C.~Feng, Q.~Chen, Y.~Wang, P.-Y. Kong, H.~Gao, and S.~Chen, ``Provision of
  contingency frequency services for virtual power plants with aggregated
  models,'' \emph{IEEE Transactions on Smart Grid}, vol.~14, no.~4, pp.
  2798--2811, 2023.

\bibitem{DER}
S.~S. Guggilam, C.~Zhao, E.~Dall’Anese, Y.~C. Chen, and S.~V. Dhople,
  ``Optimizing der participation in inertial and primary-frequency response,''
  \emph{IEEE Transactions on Power Systems}, vol.~33, no.~5, pp. 5194--5205,
  2018.

\bibitem{feng2025hybrid}
C.~Feng, L.~Huang, X.~He, Y.~Wang, F.~Dorfler, and C.~Kang, ``Hybrid
  oscillation damping and inertia management for distributed energy
  resources,'' \emph{IEEE Transactions on Power Systems}, 2025.

\bibitem{VIS}
B.~She, F.~Li, J.~Wang, H.~Cui, X.~Wang, and R.~Bo, ``Virtual inertia
  scheduling (vis) for microgrids with static and dynamic security
  constraints,'' \emph{IEEE Transactions on Sustainable Energy}, pp. 1--12,
  2024.

\bibitem{FFR3}
B.~K. Poolla, D.~Groß, and F.~Dörfler, ``Placement and implementation of
  grid-forming and grid-following virtual inertia and fast frequency
  response,'' \emph{IEEE Transactions on Power Systems}, vol.~34, no.~4, pp.
  3035--3046, 2019.

\bibitem{COM2}
Y.~Shen, W.~Wu, B.~Wang, and S.~Sun, ``Optimal allocation of virtual inertia
  and droop control for renewable energy in stochastic look-ahead power
  dispatch,'' \emph{IEEE Transactions on Sustainable Energy}, vol.~14, no.~3,
  pp. 1881--1894, 2023.

\bibitem{zhu2025optimal}
X.~Zhu, G.~Ruan, and H.~Geng, ``Optimal frequency support from virtual power
  plants: Minimal reserve and allocation,'' \emph{Applied Energy}, vol. 392, p.
  125913, 2025.

\bibitem{Zhu-TIA}
X.~Zhu, G.~Ruan, H.~Geng, H.~Liu, M.~Bai, and C.~Peng, ``Multi-objective sizing
  optimization method of microgrid considering cost and carbon emissions,''
  \emph{IEEE Transactions on Industry Applications}, vol.~60, no.~4, pp.
  5565--5576, 2024.

\bibitem{FFR2}
Y.~Zhang, S.~Wu, J.~Lin, Q.~Wu, C.~Shen, and F.~Liu, ``Frequency reserve
  allocation of large-scale res considering decision-dependent uncertainties,''
  \emph{IEEE Transactions on Sustainable Energy}, vol.~15, no.~1, pp. 339--354,
  2024.

\bibitem{DB2}
M.~Liu, F.~Bizzarri, A.~M. Brambilla, and F.~Milano, ``On the impact of the
  dead-band of power system stabilizers and frequency regulation on power
  system stability,'' \emph{IEEE Transactions on Power Systems}, vol.~34,
  no.~5, pp. 3977--3979, 2019.

\bibitem{Li-Nash}
X.~Li, G.~Ruan, and H.~Zhong, ``Low-carbon economic dispatch of bulk power
  systems using nash bargaining game,'' in \emph{2023 IEEE Power \& Energy
  Society General Meeting (PESGM)}, 2023, pp. 1--5.

\bibitem{Wei-Nash}
W.~Wei, F.~Liu, and S.~Mei, ``Nash bargain and complementarity approach based
  environmental/economic dispatch,'' \emph{IEEE Transactions on Power Systems},
  vol.~30, no.~3, pp. 1548--1549, 2015.

\end{thebibliography}
	\bibliographystyle{IEEEtran}
	
\end{document}